\documentclass[aps,prl,twocolumn,groupedaddress]{revtex4}

\usepackage{graphicx}
\usepackage{amssymb}
\usepackage{amsthm}

\usepackage[utf8]{inputenc}
\usepackage[T1]{fontenc}
\usepackage{lmodern}
\usepackage{microtype}
\usepackage[english]{babel}
\usepackage[hyperindex=true,colorlinks=true]{hyperref}
\usepackage{amsmath}
\usepackage{amsfonts}

\usepackage{url}

\def\orcid#1{\kern .08em\href{https://orcid.org/#1}{\includegraphics[keepaspectratio,width=0.7em]{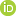}}}

\def\lnl{$N\!N\!$+$3N\text{(lnl)}$}
\def\sat{NNLO$_{\text{sat}}$}

\begin{document}

\title{\emph{Ab initio} computation of charge densities for Sn and Xe isotopes}

\author{P.\@ Arthuis \orcid{0000-0002-7073-9340}}
\email[]{p.arthuis@surrey.ac.uk}
\author{C.\@ Barbieri \orcid{0000-0001-8658-6927}}
\email[]{c.barbieri@surrey.ac.uk}
\affiliation{Department of Physics, University of Surrey, Guildford GU2 7XH, United Kingdom}
\author{M.\@ Vorabbi \orcid{0000-0002-1012-7238}}
\email[]{mvorabbi@bnl.gov}
\affiliation{
National Nuclear Data Center, Bldg. 817, Brookhaven National Laboratory, Upton, NY 11973-5000, USA}
\author{P.\@ Finelli \orcid{0000-0002-9958-993X}}
\email[]{paolo.finelli@bo.infn.it}
\affiliation{Dipartimento di Fisica e Astronomia, Università degli Studi di Bologna and INFN, Sezione di Bologna, Via Irnerio 46, I-40126 Bologna, Italy}

\date{\today}

\begin{abstract}
We present the first \emph{ab initio} calculations for open-shell nuclei past the tin isotopic line, focusing on Xe isotopes as well as doubly-magic Sn isotopes. We show that, even for moderately hard interactions, it is possible to obtain meaningful predictions and that the \sat{} chiral interaction predicts radii and charge density distributions close to the experiment.  We then make a new prediction for ${}^{100}$Sn. This paves the way for \emph{ab initio} studies of exotic charge density distributions at the limit of the present \emph{ab initio} mass domain, where experimental data is becoming available. The present study closes the gap between the largest isotopes reachable by \emph{ab initio} methods and the smallest exotic nuclei accessible to electron scattering experiments.
\end{abstract}

\maketitle


\paragraph{Introduction.}

The charge density distribution of the atomic nucleus offers a unique access to its internal structure and the spatial distribution of the nucleons, and has been probed for decades using electron scattering experiments off stable isotopes~\cite{Hahn1956, RevModPhys.28.214,Boffi1993} that have provided an impressive amount of accurate experimental data. Unfortunately, measurements on nuclei outside the valley of stability have been prevented by the difficulties associated to preparing short-lived targets, despite the interest in studying exotic nuclei presenting features like neutron halos, neutron skins or proton bubbles~\cite{Khan2008,Grasso2009,Tanihata2013,Wang2014,PhysRevC.95.034319}. Such investigations have recently been made possible with the construction of the self-confining radioactive-isotope ion target (SCRIT) at RIKEN~\cite{Suda2012,Tsukada2017,Tsukada2017a}, and will be explored as well in the next few years at FAIR by the ELISe project~\cite{Antonov2011}. By succesfully using an electron storage ring as a trap for the radioactive ions, the SCRIT experiment has been able to scatter electrons off $^{132}$Xe nuclei and recently published its first results~\cite{Tsukada2017a}. While other isotopes in the $ A \sim 130$ mass region will be studied over the next years, experimental luminosities might prevent studying lighter nuclei before future upgrades, limiting charge distribution extraction from exotic nuclei to the heavy sector.

A flourishing of new or reimplemented formalisms~\cite{Dickhoff2004,soma11a,Barbieri:2016uib,Kowalski2004,Piecuch2009,Hagen2010,Tsukiyama:2010rj,Hergert2016,Gebrerufael:2016xih,Tichai:2017rqe,Tichai:2018mll,Demol2019,Stroberg2019}, associated to new numerical approaches~\cite{Bogner2007,Bogner:2009bt,Furnstahl:2013oba} have allowed \emph{ab initio} methods to finally leave the realm of light nuclei and access mid-mass isotopes up to $ A \sim 100$~\cite{Binder:2013xaa,Morris2018} over the past few decades. But all of those approaches seem to have reached a new ceiling with the Sn isotopic line. The limitations preventing them from reaching higher masses are diverse, from interactions based on chiral Effective Field Theory ($\chi$EFT) overbinding mid-mass nuclei~\cite{Soma2014,Lapoux2016}, to numerical limitations linked to the size of the basis as well as the matrix elements storage.

Recently, new interactions have been developed~\cite{Hebeler2011,Ekstrom:2015rta,Ekstroem2018,Soma2019,Huether2019}, leading to an improvement in the reproduction of experimental data for mid-mass nuclei. New frameworks have been proposed for the treatment of both the Hamiltonian and the many-body formalism~\cite{Frame:2017fah,Tichai:2018eem,Tichai2019,Ekstroem2019a}, paving the way towards larger model spaces and promising to extend the reach of \emph{ab initio} methods within the next few years. While a first qualitative reproduction of Sn closed-shell nuclei ground-state energies had been obtained a few years ago~\cite{Binder:2013xaa}, the spectroscopy of the light end of the Sn isotopic chain has only been investigated recently~\cite{Morris2018} with an interaction able to reproduce experimental results for heavier nuclei~\cite{Hebeler2011}. This raises the question of using present day frameworks to extend the frontier of the \emph{ab initio} domain and compare with experimental charge distributions that will become available at SCRIT. Investigating discrepancies between \emph{ab initio} theoretical predictions and experimental results will allow to put new constraints on the experiment as well as to inform our theoretical models, and open the way to the study of heavy nuclei structure from first principles.

In this Letter we use self-consistent Green's function theory (SCGF)~\cite{Dickhoff2004,soma11a,Barbieri:2016uib} 
with $\chi$EFT Hamiltonians, 
present what are to our knowledge the first \emph{ab initio} calculations of charge radius, neutron skin and charge density distribution for $^{100}$Sn, $^{132}$Sn, $^{132}$Xe, $^{136}$Xe and $^{138}$Xe, and reproduce the experimental cross-section obtained at SCRIT for $^{132}$Xe.

\paragraph{Self-consistent Green's function theory.}

For solving the $A$-body Schrödinger equation, SCGF theory~\cite{Dickhoff2004,Barbieri:2016uib} expresses the nucleon dynamics in terms of one- to $A$-body propagators or Green's functions. These propagators are expanded in perturbative series which are recast into the exact Green's functions within self-consistent schemes, implicitly resuming infinite sets of diagrams. The one-body propagators are particularly interesting as they give access to all one-body observables and to the ground-state energy through the Galitskii-Migdal-Koltun sum rule~\cite{Carbone:2013eqa}. A unique and interesting feature of the one-body propagator is that it also gives access to informations on the neighbouring nuclei~\cite{Cipollone:2014hfa,Soma2020Frontiers}.

In order to obtain the one-body Green's function, one solves the intrinsically non-perturbative Dyson equation, which relies on the irreducible self-energy encoding all non-trivial many-body correlations between individual nucleons and the nuclear medium. In particular, this comprises both information on the $A$-nucleon ground state and scattering states of the $A+1$ systems, making SCGF a natural \emph{ab initio} approach for computing structure and reaction observables consistently~\cite{Waldecker2011,Idini2019}. To be able to access open-shell nuclei, where pairing has to be included for a qualitatively-correct description, Dyson SCGFs have been generalised using a $U(1)$-symmetry-breaking reference state obtained from solving the Hartree-Fock-Bogoliubov equation, yielding the Gorkov SCGF theory~\cite{soma11a}. While the broken particle-number symmetry has to be restored eventually, such a development remains to be formulated for Gorkov SCGF.

In this Letter, the self-energy is obtained via the algebraic diagrammatic construction approach, or ADC($n$)~\cite{Schirmer83,Raimondi:2017kzi}, which comprises all perturbative contributions up to order $n$ plus any infinite order resummation that is needed to preserve the spectral representation.
At the moment Dyson SCGF has been numerically implemented up to ADC(3)~\cite{Cipollone:2014hfa,Raimondi:2017kzi}, but the Gorkov formalism has only been implemented up to ADC(2)~\cite{Soma2014}, such that in the following calculations made on open-shell nuclei are done at the ADC(2) level. Both truncation levels incorporate mean-field as well as 2p1h and 2h1p contributions. While the two-body force is treated fully, the three-body force (3NF) is included into the final calculation in an effective way, as described in~\cite{Carbone:2013eqa,Cipollone:2014hfa}.

\paragraph{Results.}

For the present work we will mostly focus on the \sat{}~\cite{Ekstrom:2015rta} Hamiltonian, as it offers one of the best reproduction of radii and densities for medium-mass nuclei among chiral interactions~\cite{Soma2019}. This interaction is used bare, and incorporates two-body forces and 3NF both at next-to-next-to-leading order (NNLO) in the chiral development.
We performed calculations in a spherical harmonic oscillator basis, with frequencies ranging from $\hbar\Omega=10\text{ to }16 \text{ MeV}$, where the minimum for the ground-state energy was shown to reside by a first set of exploratory calculations. All the states of the single-particle basis up to $N_{max} = 13$ are used, i.e.~14 major shells, and one- and two-body operators are fully included. This is however not feasible for three-body operators due to the exponential increase in the number of their matrix elements and the associated storage cost, so only three-body excitations up to $E_{3max} = 16$ were considered. The restricted size of the single-particle basis and the cut on the number of three-body matrix elements prevented us from obtaining converged results for the ground-state energy, as previously observed with \sat{} on $^{78}$Ni~\cite{Hagen2016}. As such, we do not discuss such results in this Letter.

\begin{figure}
  \includegraphics[width=\linewidth]{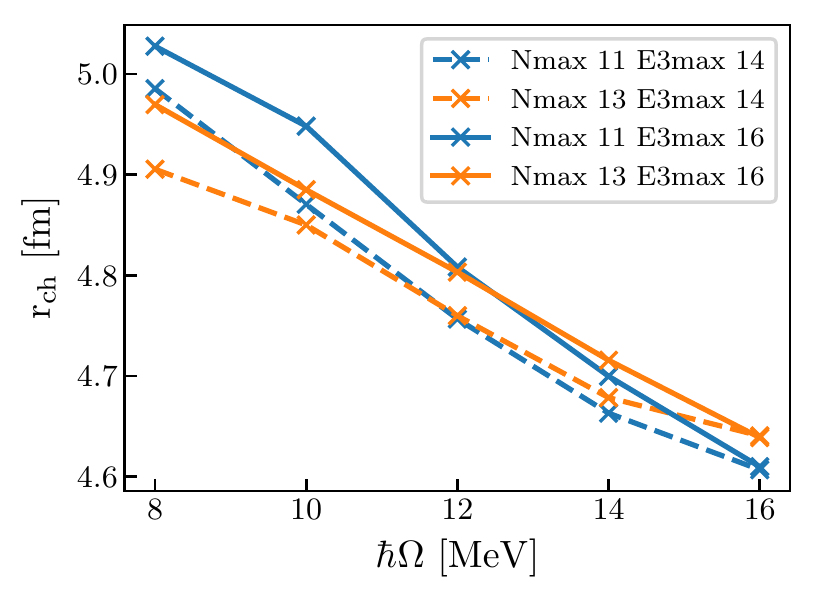}%
  \caption{\label{fig:radius_Xe132} Model space dependence of the charge radius for $^{132}$Xe, obtained from Gorkov SCGF calculations at ADC(2).}
\end{figure}

Let us first study the case of $^{132}$Xe.
Fig.~\ref{fig:radius_Xe132} provides the value of the charge radius for $^{132}$Xe obtained from Gorkov SCGF calculations at the ADC(2) truncation level over a range of harmonic oscillator frequencies that includes the optimal value. The solid (dashed) lines indicate calculations using three-body matrix elements for triplet excitations up to $E_{3max} = 16$ ($E_{3max} = 14$), while the orange (blue) lines correspond to a model space of $N_{max} = 13$ ($N_{max} = 11$). The expected behaviour, i.e.~a decrease in the radius with larger frequencies and a radius getting independent of the frequency with larger model spaces, is reproduced. The converged value of the charge radius being expected to be near the crossing of the $N_{max} = 13$ and 11 lines~\cite{Nogga2006,Bogner2008,TiMu18}, we choose here and for the other nuclei to take a conservative estimate by considering that it lies between the highest value at $\hbar\Omega =$ 10 MeV and the lowest one at 14 MeV. As can be seen here and consistently with what is obtained for the other nuclei, the cut on the three-body matrix element has only a limited effect on the value of the charge radius. Though they are not discussed here, similar results have been obtained with the other nuclei discussed in the following.

Additionally to the convergence in terms of model space and number of three-body matrix elements, the convergence in terms of the truncation scheme must be considered. Among the nuclei studied in this Letter, only  $^{100}$Sn and  $^{132}$Sn are doubly magic and can be computed at the ADC(3) truncation level. Our investigations show that, as observed previously on lighter nuclei~\cite{PhysRevC.95.034319,Barbieri2019,Soma2019}, the difference between the ADC(2) and ADC(3) values for the charge radius (and similarly for the charge density distribution) is very small, such that it is basically converged at the ADC(2) level. As such, we do not discuss differences between ADC(2) and ADC(3) results any further in this Letter. In the following, we will hence represent our results as a band obtained for frequencies from 10 to 14 MeV at $N_{max} = 13$ and from 12 to 14 MeV at $N_{max} = 11$, for $E_{3max} = 16$.

From this procedure, the charge radius of $^{132}$Xe is estimated to be $4.824 \pm 0.124$ fm, which agrees with the value extracted from the SCRIT experiment recently, namely $\langle r^2 \rangle^{1/2} = 4.79^{+0.11}_{-0.08}$ fm~\cite{Tsukada2017}. For comparison, the calculations have been reproduced using the newly-proposed \lnl{} interaction~\cite{Soma2019}, which is known to have good convergence properties with respect to the model space size and to give results similar to the very succesful 1.8/2.0(EM) interaction~\cite{Hebeler2011}. In contrast with \sat{}, the charge radius obtained for $^{132}$Xe is $4.070\pm 0.045$ fm, largely underestimating the experimental value consistently with studies on lighter nuclei~\cite{Soma2019}. Despite this failure at reproducing the experimental value, one notices that \lnl{} yields better-converged values than \sat{} as expected.

\begin{figure}
  \includegraphics[width=\linewidth]{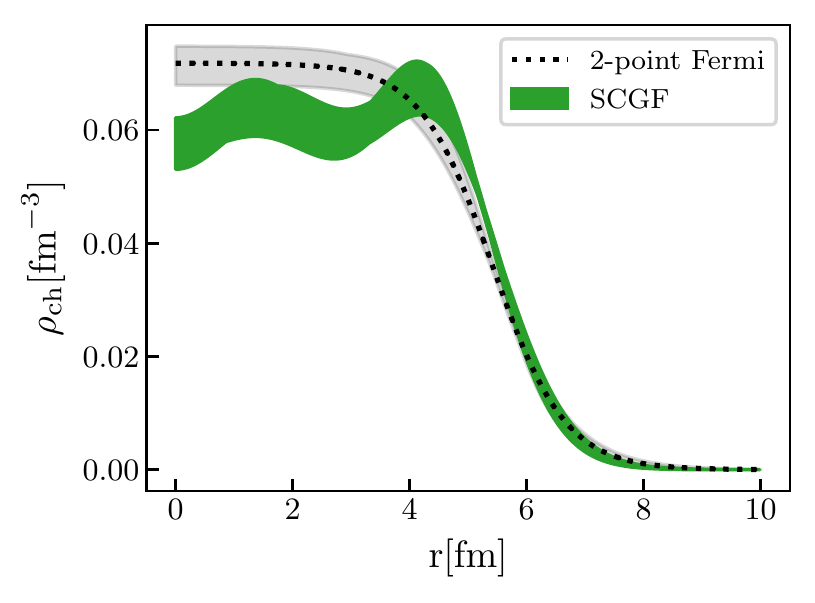}%
  \caption{\label{fig:ch_dist_Xe132} Charge density distribution for $^{132}$Xe obtained from Gorkov SCGF calculations at ADC(2). The dotted line with grey band corresponds to the two-point Fermi distribution with parameter and error bars extracted from Ref.~\cite{Tsukada2017}.}
\end{figure}

Additionally to the sole charge radius, another quantity that can be computed from SCGF calculations is the charge density distribution. In the case of $^{132}$Xe, the SCRIT group extracted the parameters $c$ and $t$ for a two-parameter Fermi charge distribution $ \rho(r) = \rho_0 / \left\lbrace 1 + \exp [4 \ln 3 (r-c) / t]\right\rbrace$. Fig.~\ref{fig:ch_dist_Xe132} displays this two-point Fermi distribution as a dotted line with a gray band representing the error bars, while the green band represents our SCGF calculations. It can be observed that while the SCGF calculations agree with the 2-point Fermi distribution at the surface of the nucleus, though slightly over-predicting the charge radius, we obtain an oscillating behaviour for the density inside the nucleus that cannot be reproduced with only a two-point Fermi distribution. Extracting a three-point Fermi distribution from the experiment would require an increase in its luminosity, such that possible discrepancies between theory and experiment cannot be discussed any further here.

\begin{figure}
  \includegraphics[width=\linewidth]{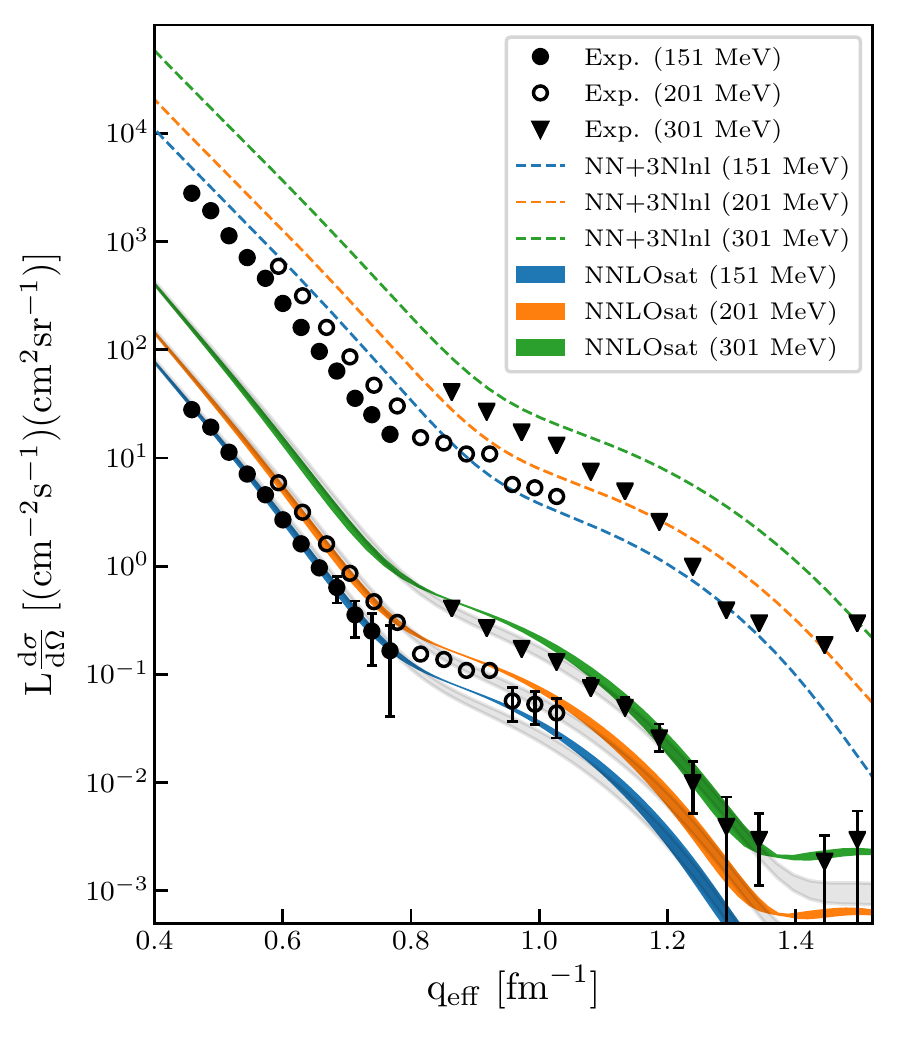}%
  \caption{\label{fig:x_sec__Xe132} Luminosity multiplied by the differential cross-section for $^{132}$Xe obtained from Gorkov SCGF calculations at ADC(2).  The values for the \lnl{} interaction have been scaled by $10^2$ for clarity. The grey bands correspond to the two-point Fermi distribution with parameter and error bars extracted from Ref.~\cite{Tsukada2017}. Experimental values are taken from~\cite{Tsukada2017}, and duplicated with a scaling of $10^2$ for comparison with \lnl{} values, where error bars have been removed for clarity.}
\end{figure}

To better gauge the discrepancies between the theoretical and experimental bands in Fig.~\ref{fig:ch_dist_Xe132}, we compare the computed electron scattering cross-sections directly to SCRIT data. Fig.~\ref{fig:x_sec__Xe132} displays the differential cross sections multiplied by the luminosity as a function of the effective momentum transfer for the three experimental electron beam energies of $E_e = 151$ MeV, 201 MeV and 301 MeV. Experimental points and error bars are taken from Ref.~\cite{Tsukada2017}. The different bands are computed using the DREPHA code \cite{drepha} starting from the nuclear charge density distributions obtained from the two-point Fermi distribution of Ref.~\cite{Tsukada2017} (grey bands) and from our SCGF calculations using \sat{} (coloured bands). 
The calculation is performed in the Distorted Wave Born Approximation (DWBA) \cite{PhysRev.95.500,Uberall:105487,CIOFIDEGLIATTI1980163}.
The results show very good agreement with the experimental values, with only an interval of effective momentum transfers between 0.8~fm$^{-1}$ and 1.1~fm$^{-1}$ being slightly off the error bars. To discard the density oscillations within the nucleus as the source of the discrepancy, we fitted a two-point Fermi density to the radius and surface predicted by the theory. Calculations using this Fermi distribution gave results within the band obtained from the genuine SCGF density. This confirms the inability of the experiment to give insights on the internal structure of the nucleus without going past the second minimum in the cross-section. As a comparison, the results obtained with the \lnl{} interaction are displayed as well, scaled upwards for clarity. Contrary to \sat{}, it fails at reproducing the experimental values, as expected with an underestimated charge radius. This demonstrates the unique capacity of \sat{} to reproduce radii and density distributions, and sets an important precedent in the use of SCGF with the \sat{} interaction for pre- or post-diction of experimental results from electron scattering off exotic nuclei. In particular, this motivates experimental measurements at higher momentum transfer to properly gauge the internal structure of nuclei.

\begin{figure}
  \includegraphics[width=\linewidth]{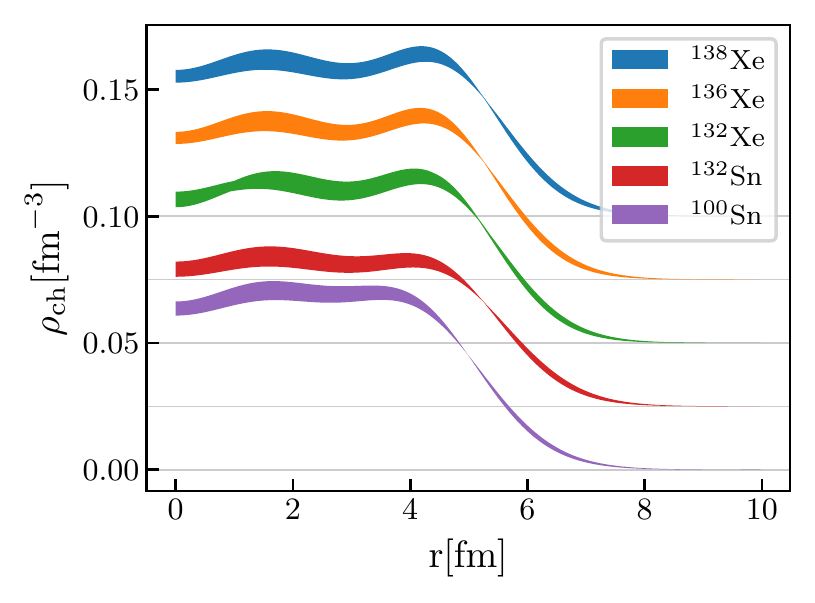}%
  \caption{\label{fig:ch_dist_all} Charge density distributions for $^{100}$Sn, $^{132}$Sn, $^{132}$Xe, $^{136}$Xe, and $^{138}$Xe obtained from Gorkov SCGF calculations. The charge density is shifted upwards by 0.025 fm$^{-3}$ between each two nuclei and the coloured bands indicate the theoretical error associated with model space convergence.}
\end{figure}

Having proved the capacity of SCGF and \sat{} to give meaningful insights on the charge radius and density distributions of $^{132}$Xe, charge densities have also been calculated for $^{100}$Sn, $^{132}$Sn, $^{136}$Xe and $^{138}$Xe for this Hamiltonian. These are displayed in Fig.~\ref{fig:ch_dist_all}. The behaviour of the charge distributions is qualitatively similar for all of them, with oscillations of the density within the nucleus and the possibility of a light depletion at its center.

\begin{table}
     \begin{tabular}{r|c|c}
 & SCGF & Exp.  \\
 \toprule
  $^{100}$Sn \ & \ 4.525 -- 4.707 \ &  \\
  $^{132}$Sn \ & \ 4.725 -- 4.956 \ & \  4.7093 \\
  $^{132}$Xe \ & \ 4.700 -- 4.948 \ & \  4.7859 \\
  $^{136}$Xe \ & \ 4.715 -- 4.928 \ & \ 4.7964 \\
  $^{138}$Xe \ & \ 4.724 -- 4.941 \ & \ 4.8279 \\
     \end{tabular}
    \caption{\label{tab:ch_rad} Charge radii in fm, obtained from SCGF calculations and \sat{}, compared with experimental values from Ref.~\cite{Angeli2013}.}
\end{table}

The charge radii extracted from our calculations are displayed for the same Sn and Xe isotopes in Tab.~\ref{tab:ch_rad} and compared with experimental results~\cite{Angeli2013}. Our results show overall a good reproduction of the experimental data and are a proof of the capacity of \sat{} to produce accurate results in the heavy nuclei regime, even despite the inability to obtain converged values for the ground-state energy. In the future, more accurate calculations with smaller errors may uncover slight differences between \sat{} and the experimental values. Among the nuclei studied, $^{100}$Sn stands out as a particularly interesting case. Sitting close to the proton dripline~\cite{Erler2012}, at the end of super-allowed $\alpha$-decay chains~\cite{Liddick2006,Seweryniak2006} and with the largest strength known in allowed $\beta$ decay~\cite{Hinke2012}, and being expected to be the heaviest doubly-magic nucleus with $N=Z$~\cite{Lewitowicz1994}, experimental data in its area are scarce~\cite{Faestermann2013}. In particular, neither its spectrum nor its radius have been measured yet. While its spectrum has recently been predicted from first principles~\cite{Morris2018}, Tab.~\ref{tab:ch_rad} displays the first \emph{ab initio} prediction of its charge radius. 

\begin{table}
     \begin{tabular}{r|c|c}
 & \sat{} & \lnl{} \\
 \toprule
  $^{100}$Sn \ & \ -0.079 -- -0.096 \ & \ -0.060 -- -0.068 \\
  $^{132}$Sn \ & \ 0.168 -- 0.197 \ & \ 0.180 -- 0.275 \\
  $^{132}$Xe \ & \ 0.103 -- 0.128 \ & \ 0.120 -- 0.152 \\
  $^{136}$Xe \ & \ 0.128 -- 0.156 \ & \ 0.134 -- 0.223 \\
  $^{138}$Xe \ & \ 0.143 -- 0.175 \ & \ 0.152 -- 0.251 \\
     \end{tabular}
    \caption{\label{tab:cn_skin} Neutron skins in fm computed with SCGF. Each interval indicates the theoretical error associated with model space convergence.}
\end{table}

Neutron skins are directly related to the density dependence of the nuclear symmetry energy. SCGF calculations in the mass range $ A = 40-64$~\cite{Soma2019} suggest that \sat{} and \lnl{} yield nearly identical skins, in spite of their differences in the prediction of radii~\cite{SomaPrivComm}. These neutron skins tend to be systematically higher (or smaller proton skins) than the experimental findings from Ref.~\cite{Trzcinska2001} but are within the reported error bars. Our results for Sn and Xe are shown in Table~\ref{tab:cn_skin} for both Hamiltonians. Although they are consistent with each other within the uncertainties from the model space convergence, \lnl{} gives slightly higher values. These differences correlate with the differences in charge radii as found in Ref.~\cite{Hagen2016Ca48}.   For $^{132}$Sn, neutron skins of 0.24(4)~fm~\cite{Klimkiewicz2007} and 0.258(24)~fm~\cite{Carbone2010skinSn132} have been extracted from measurements of low lying dipole excitations, while Skyrme functionals predict 0.263-0.294~fm~\cite{Burrello2019}. The \sat{} is in disagreement with these values as can be expected since it is already known to miss the expected symmetry energy at saturation density~\cite{Carbone2020Tdep}.
These results stress the need for accurate experimental data in the neutron-rich areas of the nuclear chart, where \emph{ab initio} calculations tend to struggle to reproduce radii~\cite{Ruiz2016}.

\paragraph{Conclusions.}

Our calculations demonstrated the capacity of SCGF and the \sat{} interaction to give a meaningful estimation of the charge radius and charge density distribution of heavy nuclei up to mass $A = 138$ which had never been studied before.
We computed successfully the charge radius, density distribution and neutron skins of $^{132}$Sn, $^{132}$Xe, $^{136}$Xe, and $^{138}$Xe, mostly agreeing with known experimental values, and gave the first \emph{ab initio} prediction for the charge radius and density distribution of $^{100}$Sn. In particular, we reproduced the experimental cross-section of the SCRIT electron scattering experiment for $^{132}$Xe, demonstrating the capacity of \emph{ab initio} methods with well-designed chiral interactions to be used for the internal structure study of heavy exotic nuclei, alongside new experimental facilities. Our errors bars, though conservative, are small enough to shed light on discrepancies with experimental values, informing theory and putting constraints on experiments. In particular, our results are a motivation for measurements at higher momentum transfer to probe the internal structure of the nuclei.


\begin{acknowledgments}
The authors are grateful to Petr Navrátil for providing matrix elements of the \sat{} and \lnl{} interactions and to him, Carlotta Giusti and Vittorio Somà for several useful discussions. 
This work is supported by the UK Science and Technology Facilities Council (STFC) through grants ST/P005314/1 and ST/L005516/1.
Calculations were performed by using HPC resources at the DiRAC DiAL system at the University of Leicester, UK, (BIS National E-infrastructure Capital Grant No. ST/K000373/1 and STFC Grant No. ST/K0003259/1).
The work at Brookhaven National Laboratory was sponsored by the Office of Nuclear Physics, Office of Science of the U.S. Department of Energy under
Contract No. DE-AC02-98CH10886 with Brookhaven Science Associates, LLC.
\end{acknowledgments}

\bibliography{references}

\end{document}